\documentclass[a4paper]{JHEP3}
\title{$U(N|M)$ quantum mechanics on K\"ahler manifolds}
\author{Fiorenzo Bastianelli and Roberto Bonezzi\\ 
Dipartimento di Fisica, Universit\`{a} di Bologna and INFN, Sezione di Bologna\\
via Irnerio 46, I-40126 Bologna, Italy\\ 
E-mail: \email{bastianelli@bo.infn.it, bonezzi@bo.infn.it}} 
\abstract{
We study the extended supersymmetric quantum mechanics, with supercharges
transforming in the fundamental representation of  $U(N|M)$,
as realized in certain one-dimensional nonlinear sigma models
with K\"ahler manifolds as target space.  
We discuss the symmetry algebra characterizing these models and,
using operatorial methods, compute the heat kernel 
in the limit of short propagation time. 
These models are relevant for studying the quantum properties of a certain class 
of higher spin field equations in first quantization.}

\keywords{Sigma models, Extended supersymmetry}
\usepackage[latin1]{inputenc}
\usepackage{indentfirst}
\usepackage{graphicx}
\usepackage{amssymb}
\usepackage{amsmath}
\usepackage{latexsym}

\newcommand{\be}{\begin{equation}}
\newcommand{\ee}{\end{equation}}
\newcommand{\bea}{\begin{eqnarray}}
\newcommand{\eea}{\end{eqnarray}}

\newcommand{\ph}[1]{\phantom{#1}}
\newcommand{\de}{\partial} 
\newcommand{\braket}[2]{ \langle #1 \lvert #2 \rangle }  
\newcommand{\ket}[1]{\lvert #1 \rangle} 
\newcommand{\bra}[1]{\langle #1 \rvert} 
\newcommand{\Braket}[3]{\langle #1 \lvert #3 \rvert #2 \rangle }  

\newcommand{\abs}[1]{\left\lvert #1 \right\rvert}

\begin{document}


\section{Introduction}

$O(N)$ spinning particles \cite{Gershun:1979fb,Howe:1988ft,Kuzenko:1995mg}
have been useful to describe higher spin  fields in first quantization 
\cite{Siegel:1999ew,Bastianelli:2007pv}.
Similarly,  $U(N)$ spinning particles \cite{Marcus:1994em, Marcus:1994mm}
have been instrumental to discover 
a new class of higher spin field equations which possess a novel type of gauge 
invariance \cite{Bastianelli:2009vj}.
To investigate  the quantum properties of these equations in their worldline
formulation, it is important to study the related quantum mechanics.
It is the purpose of this paper to discuss these quantum mechanics,
which in the most general case take the form of nonlinear sigma models. 

First we shall discuss linear sigma models, i.e.
models with flat complex space $\mathbb{C}^d$ as target space.
These sigma models exhibit a $U(N)$  extended supersymmetry on the worldline.
They define ``spinning particle'' models once the 
extended supersymmetry is made local.
It is useful, and almost effortless, to  extend these models by adding
extra bosonic coordinates. This extension produces $U(N|M)$ sigma models,
by which we mean  sigma models with a worldline extended supersymmetry
characterized by supercharges transforming in the fundamental representation 
of  $U(N|M)$ (i.e. $U(N|M)$ is the $R$-symmetry group of the supersymmetry algebra). 
This extension may be useful for constructing wider classes
of spinning particles, as happened in the case of the $OSp(N|2M)$ extension
\cite{Hallowell:2007qk} of the standard $O(N)$ supersymmetric quantum mechanics,
used for example in \cite{Campoleoni:2008jq,Alkalaev:2008gi,Bastianelli:2009eh,Cherney:2009vg}
to describe higher spin fields.   We present these quantum mechanical models 
and their symmetry algebra in section 2.

In section 3 we consider sigma models with generic K\"ahler 
manifolds as target spaces.  The symmetry algebra
gets modified by the geometry, so that it will not be always possible to
gauge the extended supersymmetry to obtain spinning particles 
and corresponding higher spin equations.
This signals the difficulties of coupling higher spin fields to generic
backgrounds, not to mention the even more difficult problem of constructing nonlinear 
field equations. 
However, on special backgrounds one can find a 
deformed $U(N|M)$ susy algebra that becomes first class, so that it can be gauged
to produce consistent spinning particles. An example is the case of K\"ahler manifolds 
with constant holomorphic sectional curvature. No restrictions apply to the special cases  of 
$U(1|0)$ and $U(2|0)$, whose susy algebra can be gauged 
to produce nontrivial field equations on any K\"ahler space, in analogy
with standard $N=1$ and $N=2$ susy quantum mechanics on arbitrary riemannian manifolds
(i.e. $O(1)$ and $O(2)$ quantum mechanics in the language used above).

Nevertheless, before gauging, the $U(N|M)$ quantum mechanics 
here constructed are perfectly consistent on any K\"ahler 
manifold, and even posses conserved supercharges
when the Riemann tensor obeys a 
locally symmetric space condition (again in close analogy
with the riemannian case \cite{Hallowell:2007qk}).
Thus, in section 4 we work with an arbitrary K\"ahler manifold and
 compute the quantum mechanical 
transition amplitude in euclidean time
(i.e. the heat kernel)  in the limit of short propagation time and 
using operatorial methods. 
This last result is going to be particularly useful for obtaining an unambiguous
construction of the corresponding path integral, which is needed when 
considering worldline applications.
This is indeed one of our future aims, namely using worldline 
descriptions of higher spin fields to obtain useful and computable
representations of their one-loop effective actions, as done in \cite{Ba:2005vk} 
for the $O(2)$ spinning particle. In that case a worldline
representation allowed to compute in a single stroke the first few heat kernel 
coefficients and prove various duality relations for massless and massive $p$-forms 
in arbitrary dimensions. 
Finally, we present our conclusions and outlook in section 5, and confine 
to the appendices details of our calculations.

\section{Linear $U(N|M)$ sigma model}

We introduce here the $U(N|M)$ extended supersymmetric quantum mechanics.
In the most simple case it describes the motion of a particle in $\mathbb{C}^d$,  
the flat complex space of $d$ complex dimensions with coordinates
($x^\mu$, $\bar x^{\bar\mu}$), $\mu=1,...,d$.
The flat metric in these complex coordinates is simply
$\delta_{\mu\bar\nu}$, and we use it to raise and lower indices.
In addition, the particle carries extra degrees of freedom
described by worldline Dirac fermions ($\psi^\mu_a$, $\bar\psi^a_\mu$)
and complex bosons ($z^\mu_\alpha$, $\bar z^\alpha_\mu$),
where $a=1,...,N$ and $\alpha=1,...,M$ are indices in the $U(N)$ and $U(M)$ subgroups 
of $U(N|M)$, respectively.  These extra degrees of freedom can be interpreted 
as worldline superpartners of the coordinates ($x^\mu$, $\bar x^{\bar\mu}$). 
Of course, when the superpartners have bosonic character one finds a kind of ``bosonic'' supersymmetry, 
that generalizes usual concepts.
With these degrees of freedom at hand 
the phase space lagrangian defining our model
has the standard form ${\cal L}\sim p\dot q - H$, namely
\begin{equation}\label{U(N|M) flat lagrangian}
\mathcal{L}=p_\mu\dot x^\mu+\bar p_{\bar\mu}\dot{\bar
x}^{\bar\mu}+i\bar\psi_\mu^a\dot\psi^\mu_a+i\bar z_\mu^\alpha\dot z^\mu_\alpha-p_\mu\bar p^\mu\;.
\end{equation}
This model enjoys a $U(N|M)$ extended supersymmetry, which we are going 
to describe directly in the quantum case.

The fundamental (anti)-commutators are easily read off from \eqref{U(N|M) flat lagrangian}
\begin{equation}\label{CCR}
\begin{split}
& [x^\mu, p_\nu]=i\hbar\delta^\mu_\nu\;,\qquad\quad\;
 [\bar x^{\bar\mu}, \bar p_{\bar\nu}]=i\hbar\delta^{\bar\mu}_{\bar\nu}\\[1.2mm]
& \{\psi_a^\mu, \bar\psi^b_\nu\}=\hbar\delta_a^b\delta^\mu_\nu\;, \qquad
 [z_\alpha^\mu, \bar z^\beta_\nu]=\hbar\delta_\alpha^\beta\delta^\mu_\nu \; .
\end{split}
\end{equation}

The $U(N|M)$ charges are readily constructed from the worldline operators
\begin{equation}
\begin{split}
J^a_b &= \frac12[\bar\psi^a_\mu, \psi^\mu_b] - c\hbar \delta^a_b
=\bar\psi^a_\mu \psi^\mu_b -m\hbar\delta^a_b
\qquad\text{$U(N)$ subgroup},\\
J^\alpha_\beta &= \frac12\{\bar z^\alpha_\mu, z_\beta^\mu\} + c\hbar \delta^a_b
= \bar z^\alpha_\mu z_\beta^\mu + m \hbar\delta^\alpha_\beta
\qquad\text{$U(M)$ subgroup},\\
J_b^\alpha &= \bar z^\alpha_\mu\psi_b^\mu\;,\qquad \ \; J^a_\beta=\bar\psi^a_\mu z_\beta^\mu\qquad\qquad\qquad\  \text{$U(N|M)$ fermionic
generators,}
\end{split}
\end{equation}
where $m= c+\frac{d}{2}$. They obey the $U(N|M)$ algebra
\begin{equation}\label{U(N|M) internal superalgebra flat}
\begin{split}
[J^a_b, J^c_d] &= \hbar\,(\delta^c_bJ^a_d-\delta^a_dJ^c_b)\\
[J^\alpha_\beta, J^\gamma_\delta] &= \hbar\,(\delta^\gamma_\beta J^\alpha_\delta-\delta^\alpha_\delta J^\gamma_\beta)\\
[J^a_b, J^\alpha_c] &= -\hbar\,\delta^a_c J^\alpha_b\;,\quad [J^a_b, J^c_\alpha]=\hbar\,\delta^c_b J^a_\alpha\\
[J^\alpha_\beta, J^\gamma_a] &= \hbar\,\delta^\gamma_\beta J^\alpha_a\;,\quad [J^\alpha_\beta, J^a_\gamma]=-\hbar\,\delta^\alpha_\gamma J^a_\beta\\
\{J_a^\alpha, J_\beta^b\} &=\hbar\,(\delta_a^bJ^\alpha_\beta+\delta^\alpha_\beta J_a^b)\;.
\end{split}
\end{equation}
In the definition of these charges we have used a ``graded symmetric'' ordering prescription modified 
by an arbitrary central charge $c$ that specifies possible different orderings allowed by the symmetry algebra.
The possibility of inserting the central charge is related to the 
algebraic fact that $U(N|M) = U(1)\times SU(N|M)$.
 All these charges commute with the hamiltonian
$H=p_\mu\bar p^\mu$ and are conserved.

Other conserved quantities are the 
supersymmetric charges involving the space momenta: there are 
$2N$ fermionic supercharges $Q_a=\psi_a^\mu\, p_\mu$, $\bar Q^a=\bar\psi^a_\mu\, \bar p^\mu$, 
and $2M$ bosonic charges $Q_\alpha=z_\alpha^\mu\, p_\mu$, $\bar Q^\alpha=\bar z^\alpha_\mu\, \bar p^\mu$. 
All these operators form the $U(N|M)$ extended superalgebra that, together with the 
$U(N|M)$ internal algebra \eqref{U(N|M) internal superalgebra flat}, is given by the following relations
\begin{equation}
\begin{split}
[J^a_b, Q_c] &= -\hbar\,\delta^a_c\,Q_b\;,\quad\; [J^a_b, \bar Q^c] = \hbar\,\delta^c_b\,\bar Q^a \\
[J^\alpha_\beta,Q_\gamma] &= -\hbar\,\delta^\alpha_\gamma\,Q_\beta\;,\quad [J^\alpha_\beta, \bar Q^\gamma] =\hbar\,\delta^\gamma_\beta\,\bar Q^\alpha \\
[J^\alpha_a, Q_\beta] &=-\hbar\,\delta^\alpha_\beta\,Q_a\;,\quad [J_\alpha^b, \bar Q^\beta] =\hbar\,\delta_\alpha^\beta\,\bar Q^b\\
\{J^a_\alpha, Q_b\} &= \hbar\,\delta^a_b\,Q_\alpha\;,\quad\ \ \{J^\alpha_b, \bar Q^c\}=\hbar\,\delta^c_b\,\bar Q^\alpha\\[2mm]
\{Q_a, \bar Q^b\}&=\hbar\,\delta_a^b\,H\;,\qquad[Q_\alpha, \bar Q^\beta]=\hbar\,\delta_\alpha^\beta\,H\;.
\end{split}
\end{equation}
(Anti)-commutators needed to close the algebra and not explicitly reported vanish.

All these relations can be written in a more covariant way.
In order to show up the full supergroup structure, let us introduce the superindex $A=(a,\alpha)$
and the $U(N|M)$ metrics
\begin{equation}
\delta^A_B=\left(\begin{array}{cc}\delta^a_b & 0\\
0 &\delta^\alpha_\beta \end{array}\right)\;, \qquad
\epsilon^A_B=\left(\begin{array}{cc}-\delta^a_b & 0\\
0 &\delta^\alpha_\beta \end{array}\right)\;.
\end{equation}
The internal fermions and bosons are grouped into the fundamental and anti-fundamental
representations of the supergroup, $Z_A^\mu=(\psi_a^\mu, z_\alpha^\mu)$, $\bar
Z^A_\mu=(\bar\psi^a_\mu, \bar z^\alpha_\mu)$. The fundamental (anti)-commutation relations
can be written as $[Z^\mu_A, \bar Z_\nu^B\}=\hbar\,\delta_A^B\, \delta^\mu_\nu $,
or equivalently as $[\bar Z_\nu^B,Z^\mu_A,\}=-\hbar\,\epsilon_A^B\, \delta^\mu_\nu $.
Here the graded
commutator is used: $[A,B\}$ is defined as anti-commutator for $A$ and $B$ both fermionic, and as a
commutator otherwise. 
Then we collect all the $U(N|M)$ generators in
\begin{equation}
J^A_B=\left(\begin{array}{cc}J^a_b & J^a_\beta \\ J^\alpha_b & J^\alpha_\beta \end{array}\right)=
 \bar Z^A_\mu Z_B^\mu + m\hbar\, \epsilon^A_B \;.
\end{equation}
With these notations at hand the entire
superalgebra \eqref{U(N|M) internal superalgebra flat} is packaged into the single relation
\begin{equation}\label{U(N|M) supergroup notation}
[J^A_B, J^C_D\}=\hbar\,(\delta^C_B\,
J^A_D\pm\delta^A_D\, J^C_B)\;,
\end{equation}
where the plus sign refers to the case with $J^A_B$ and $J^C_D$ both fermionic,
and the minus sign to the other possibilities. 

By means of this supergroup notation, the supercharges are written as $Q_A=(Q_a,
Q_\alpha)$ and $\bar Q^A=(\bar Q^a, \bar Q^\alpha)$, and the above superalgebra 
is summarized by
\begin{equation}\label{U(N|M) complete superalgebra flat}
\begin{split}
[J^A_B, Q_C\} &= \pm\hbar\,\delta^A_C\,Q_B\; , \qquad
[J^A_B, \bar Q^C\} = \hbar\,\delta^C_B\,\bar Q^A\\
[Q_A, \bar Q^B\} &= \hbar\,\delta_A^B\,H\;,
\end{split}
\end{equation}
where $\pm$ stands for plus for $J^A_B$ and $Q_C$ both fermionic, and minus
otherwise.

All these quantum mechanical operators have simple geometrical meanings 
in terms of differential operators living on $\mathbb{C}^d$.
Let us  give a brief description. 
Generic wave functions of the Hilbert space can be represented by 
functions of the coordinates $(x,\bar x, \psi, z)$.
Expanding them in $\psi^\mu$ and $z^\mu$ shows how they contain
all possible tensors with $N+M$ blocks of holomorphic indices.
Each of the first $N$ blocks of indices is totally antisymmetric, 
while each of the last $M$ blocks of indices is totally symmetric.
In formulae
\begin{equation}\label{general expansion}
\begin{split}
\phi(x,\bar x, \psi, z) \sim & \sum_{A_i=0}^d \sum_{B_i=0}^\infty
\phi_{[\mu_1^1..\,\mu_{A_1}^1],\ldots,\,[\mu_1^N ..\,\mu_{A_N}^N],
(\nu_1^1..\,\nu_{B_1}^1),\ldots,\,(\nu_1^M ..\,\nu_{B_M}^M)} (x,\bar{x})\\
& \times \Big (\psi_1^{\mu_1^1}..\,\psi_1^{\mu_{A_1}^1}\Big )
\ldots
\Big (\psi_N^{\mu_1^N}..\,\psi_N^{\mu_{A_N}^N}\Big )
\Big (z_1^{\nu_1^1}..\,z_1^{\nu_{B_1}^1}\Big )
\ldots
\Big (z_M^{\nu_1^M}..\, z_M^{\mu_{B_M}^M}\Big ) \; .
\end{split}
\end{equation}
The quantum mechanical operators take the form of differential 
operators acting on these tensors.
The hamiltonian is proportional the standard laplacian
$H\sim  \de_\mu\bar\de^{\mu} = \delta^{\mu\bar{\nu}}\,\de_\mu\bar\de_{\bar{\nu}}$.
The supercharge $Q_a$ acts as the Dolbeault operator $\de$ 
restricted to the antisymmetric indices of block ``$a$", and 
$\bar Q^a$ as its adjoint $\de^\dagger$. Similarly the ``bosonic'' 
supercharge  $Q_\alpha$ is realized as a symmetrized gradient acting on the 
symmetric indices of block ``$\alpha$", and 
 $\bar Q^\alpha$ is its adjoint, taking the form of a divergence.
 The action of the $U(N|M)$ operators, i.e. the $J^A_B$ charges,
 is also amusing: they perform certain (anti)-symmetrizations on the tensors
 indices, and we leave it to the interested reader to work them out  explicitly.
The algebra of these differential/algebraic operators, as encoded in the susy algebra, 
is only valid in flat space. In the next section we will see
how this algebra extends to generic K\"ahler manifolds.

\section{Nonlinear $U(N|M)$ sigma model}

We now extend the previous construction to nonlinear sigma models
with generic K\"{a}hler manifolds as target spaces. 
On K\"{a}hler manifolds,  in holomorphic coordinates,
the only non vanishing components of the metric 
are $g_{\mu\bar\nu}=g_{\bar\nu\mu}$, 
and similarly 
$\Gamma^\mu_{\nu\lambda}$ and $\Gamma^{\bar\mu}_{\bar\nu\bar\lambda}$ 
are the only non vanishing components of the connection.
We use the following conventions for curvatures
\begin{equation}
 R^\mu_{\ph\mu\nu\bar\sigma\lambda}=\de_{\bar\sigma}\Gamma^\mu_{\nu\lambda}\;,\quad R^\mu_\nu=-g^{\bar\sigma\lambda}\,R^\mu_{\ph\mu\nu\bar\sigma\lambda}\;,\quad R=R^\mu_\mu \;,
\end{equation}
and denote by $g=\det(g_{\mu\bar\nu})$ the determinant of the metric,
 as standard in K\"ahler geometry.

The classical phase space lagrangian with a minimally covariantized 
hamiltonian becomes
\begin{equation}\label{U(N|M) curved phase space lagrangian}
\mathcal{L}=p_\mu\dot x^\mu+\bar p_{\bar\mu}\dot{\bar x}^{\bar\mu}
+i\bar Z_\mu^A \dot Z^\mu_A
-g^{\mu\bar \nu} (p_\mu -i\Gamma^\lambda_{\mu\sigma}{\bar Z^A_\lambda}  
Z^\sigma_A )  
 \bar p_{\bar \nu}
\end{equation}
though, for future applications, it will be useful to consider more general hamiltonians. 
The corresponding configuration space lagrangian is
the typical one for nonlinear sigma models
\begin{equation}\label{U(N|M) curved conf space lagrangian}
\mathcal{L}=g_{\mu\bar \nu} \dot x^\mu \dot{\bar x}^{\bar\nu}
+i\bar Z_\mu^A \frac{D Z^\mu_A}{dt}
\end{equation}
where the covariant time derivative is given by 
$ \frac{DZ^\mu_A}{d t}=\dot Z^\mu_A+\dot x^\nu\,\Gamma^\mu_{\nu\sigma}
\,Z^\sigma_A $.

In the quantum case, it will be crucial to resolve ordering ambiguities
by demanding target space covariance.
Before discussing the quantum operators, let us 
make a few comments.
We treat the $\bar Z_\mu^A$ fields as momenta, as such they have a natural 
lower holomorphic curved index. In this situation there
is no real advantage in introducing a vielbein, so we will avoid introducing one.
Also, the holonomy group of a K\"{a}hler manifold of complex dimensions 
$d$ is $U(d)$, and it will  be convenient to define the $U(d)$ generators
\begin{equation}
M_\mu^\nu=\frac12[\bar\psi^a_\mu, \psi_a^\nu]+\frac12\{\bar z^\alpha_\mu, z^\nu_\alpha\} -k\hbar\delta^\nu_\mu 
\end{equation} 
where $k$ is  a central charge parametrizing different orderings allowed by the $U(d)=U(1)\times SU(d)$ symmetry.
These generators can be written as well as 
 \begin{equation}\label{s charge}
 M_\mu^\nu=  \bar Z^A_\mu Z^\nu_A -s\hbar\delta^\nu_\mu 
\end{equation} 
with $s=k+\frac{N-M}{2}$.
They satisfy the correct $U(d)$ algebra
\begin{equation}
[M^\mu_\nu,M^\rho_\sigma]=\hbar\,\delta^\mu_\sigma\, M^\rho_\nu-\hbar\,\delta^\rho_\nu\, M^\mu_\sigma\;.
\end{equation}

We are now ready to discuss the covariantization of the quantum operators
belonging to the $U(N|M)$ extended supersymmetry algebra.
As we shall see, not all of the charges generate symmetries 
on generic K\"ahler manifolds:  some of them do
not commute with the hamiltonian and thus are not conserved.

It is easiest to start with the generators of $U(N|M)$. They are left unchanged 
as the metric does not enter their definition: 
$J^A_B=\bar Z^A_\mu Z_B^\mu+m\hbar \epsilon^A_B$.
They satisfy the same $U(N|M)$
symmetry algebra given in eq. \eqref{U(N|M) supergroup notation}.

Now we consider the $Q$ supercharges. To covariantize them
we introduce covariant momenta
\begin{equation}
\bar\pi_{\bar\mu}=g^{1/2}\,\bar p_{\bar\mu}\,g^{-1/2}\;, \qquad
\pi_\mu=g^{1/2}\,\big(p_\mu-i\,\Gamma^\lambda_{\mu\sigma}\,M^\sigma_\lambda\big)\,g^{-1/2}\;,
\end{equation}
and write down covariantized supercharges as
\begin{equation}
Q_A=Z_A^\mu\,\pi_\mu\;,\qquad\bar Q^A=\bar Z_\mu^A\,g^{\mu\bar\nu}\,\bar\pi_{\bar\nu}\;.
\end{equation}
Similarly,  the covariant hamiltonian operator is given by
\begin{equation}\label{H_0}
H_0=g^{\bar\mu\nu}\bar\pi_{\bar\mu}\pi_\nu=g^{1/2}\, g^{\bar\mu\nu} \bar
p_{\bar\mu}\,\big(p_\nu-i\,\Gamma^\lambda_{\nu\sigma}\,M^\sigma_\lambda\big)\,g^{-1/2}\;.
\end{equation}

At this stage it is worthwhile to spend some words on the hermiticity properties of our operators: 
since the  $\bar Z^A_\mu$ fields are defined as independent variables 
with lower holomorphic indices,
but hermitian conjugation of vector indices naturally sends holomorphic
into anti-holomorphic indices, and vice versa, 
the natural definition of the adjoint of $Z_A^\mu$ is
$(Z_A^\mu)^\dagger=\bar Z_\nu^A\,g^{\nu\bar\mu}$. In this way, hermitian conjugation of the
momentum is nontrivial: if $[p_\mu,Z_A^\nu]=0$, it must hold that
$[(p_\mu)^\dagger,(Z^\nu_A)^\dagger]=[(p_\mu)^\dagger, \bar Z^A_\lambda\,g^{\lambda\bar\nu}]=0$ as well. Requiring
this property we find
\begin{equation}
\big(p_\mu\big)^\dagger=\bar
p_{\bar\mu}-i\,\Gamma^{\bar\lambda}_{\bar\mu\bar\sigma}\,M^\lambda_\sigma\,g^{\sigma\bar\sigma}g_{\lambda\bar\lambda}\;.
\end{equation}
Now, if we define the supercharges in the natural way written above, namely
 $Q_A=Z_A^\mu\,\pi_\mu$ and $\bar Q^A=\bar
Z^A_\mu\,g^{\mu\bar\nu}\,\bar\pi_{\bar\nu}$, then it results that
$(Q_A)^\dagger=\bar Q^A$ and $H_0^\dagger=H_0$. Note that
the power of the metric determinant entering the various operators
is necessary for verifying the hermiticity properties.

Let us now consider their algebra.
The first line of \eqref{U(N|M) complete superalgebra flat} simply
states that $Q_A$ and $\bar Q^A$ belong to the fundamental and anti-fundamental
representation of $U(N|M)$, and one can check that these relations remain unchanged even in curved space,
\begin{equation}\label{JQ}
[J^A_B, Q_C\} = \pm\hbar\,\delta^A_C\,Q_B\; , \qquad
[J^A_B, \bar Q^C\} = \hbar\,\delta^C_B\,\bar Q^A \;.
\end{equation}
On the other hand the last relation becomes
\begin{equation}\label{QbarQ curved space}
[Q_A, \bar Q^B\}=\hbar\,\delta_A^B\,H_0+\hbar\,Z_A^\mu\bar Z^B_\nu\,R^{\nu\ph\mu\lambda}_{\ph\nu\mu\ph\lambda\sigma}M^\sigma_\lambda\; .
\end{equation}
The minimal covariant hamiltonian $H_0$, emerging 
from this commutator as the term multiplying $\delta_A^B$
and already given in \eqref{H_0},  does not conserve the supercharges
except than in flat space; in fact the commutator between $H_0$ and $Q$ 
does not vanish and reads
\begin{equation}\label{[Q,H_0], U(N|M)}
\begin{split}
[Q_A, H_0]&=\hbar\,Z_A^\mu\,R^{\nu\ph\mu\lambda}_{\ph\nu\mu\ph\lambda\sigma}\,M_\lambda^\sigma\,\pi_\nu
+\hbar^2\,Z_A^\mu\,R^\nu_\mu\,\pi_\nu\\
[\bar Q^A, H_0]&\equiv-[Q_A, H_0]^\dagger\;.
\end{split}
\end{equation}
$H_0$ is a central operator only in flat space.  
Finally, it is simple to verify that 
\begin{equation}\label{QQ}
[Q_A, Q_B\}= [\bar Q^A, \bar Q^B\}= 0  \; .
\end{equation}
Relations \eqref{JQ}, \eqref{QbarQ curved space}, \eqref{[Q,H_0], U(N|M)}
and \eqref{QQ}, together with \eqref{U(N|M) supergroup notation},
describe the deformation of the $U(N|M)$ supersymmetry algebra
realized by our quantum nonlinear sigma model on a K\"ahler manifold.
Supersymmetry is broken as the supercharges are not conserved. 
Only on flat spaces the hamiltonian $H_0$ becomes central and the supercharges 
get conserved.

Given this state of affairs,
one may try to redefine the hamiltonian in an attempt to make it central
on more general backgrounds, thus recovering conserved supercharges.
For this purpose, we add to $H_0$ several non minimal couplings
\begin{equation}\label{H with cs}
H=H_0+c_1R^{\nu\ph\mu\lambda}_{\ph\nu\mu\ph\lambda\sigma}\,M^\mu_\nu\,M^\sigma_\lambda+c_2\,\hbar\,R^\mu_\nu\,M^\nu_\mu+c_3\,\hbar^2R\;.
\end{equation}
With these generic couplings \eqref{[Q,H_0], U(N|M)} becomes
\begin{equation}\label{[Q,H], U(N|M) generic couplings}
\begin{split}
[Q_A, H]&=\hbar\,(1+2c_1)\,Z_A^\mu\,R^{\nu\ph\mu\lambda}_{\ph\nu\mu\ph\lambda\sigma}\,M^\sigma_\lambda\,\pi_\nu+\hbar^2\,(1+c_1+c_2)\,Z^\mu_A\,R^\nu_\mu\,\pi_\nu\\
&-i\hbar c_1Z^\rho_A\nabla_\rho R^{\nu\ph\mu\lambda}_{\ph\nu\mu\ph\lambda\sigma}\,M^\mu_\nu\,M^\sigma_\lambda-i\hbar^2
c_2\,Z^\sigma_A\nabla_\sigma R^\mu_\nu\,M^\nu_\mu-i\hbar^3 c_3\,Z^\mu_A\nabla_\mu R\;.
\end{split}
\end{equation}
We see that for the choice $c_1=-\frac12$, $c_2=-\frac12$ and generic $c_3$, 
the terms in the first line
proportional to the covariant momentum $\pi_\nu$ vanish and, choosing $c_3=0$
for simplicity, we identify a canonical hamiltonian $H_{(c)}$
so that eq. \eqref{[Q,H], U(N|M) generic couplings} reduces to
\begin{equation}\label{QH commutator}
[Q_A,H_{(c)}]=\frac{i\hbar}{2}\,Z^\rho_A\,\nabla_\rho R^{\nu\ph\mu\lambda}_{\ph\nu\mu\ph\lambda\sigma}\,M^\mu_\nu\,M^\sigma_\lambda+ \frac{i\hbar^2}{2}\,Z^\sigma_A\,\nabla_\sigma R^\mu_\nu\,M^\nu_\mu\;,
\end{equation}
showing that $H_{(c)}$ is central on locally symmetric spaces. Of course, also the graded commutator
\eqref{QbarQ curved space} changes and becomes
\begin{equation}\label{QbarQ anticomm}
[Q_A, \bar Q^B\}=\hbar\,\delta_A^B\,H_{(c)}+\hbar\,R^{\nu\ph\mu\lambda}_{\ph\nu\mu\ph\lambda\sigma}\left(Z^\mu_A\bar
Z_\nu^B+\frac12\,\delta_A^B\,M^\mu_\nu\right)M^\sigma_\lambda+\frac12\,\hbar^2\,\delta_A^B\,R^\mu_\nu\,M^\nu_\mu\;.
\end{equation}
Thus one concludes that with the redefinition of the hamiltonian given above
the supercharges are conserved on locally symmetric K\"ahler manifolds. 

One of the most interesting applications of the nonlinear sigma models 
discussed so far is to use them to construct spinning particles and related higher spin 
equations. This is achieved
 by gauging the extended susy algebra 
identified by the charges $(H, Q_A, \bar Q^A, J_A^B)$,  possibly with a suitable
redefinition of the hamiltonian.
Unfortunately, we see that on generic K\"ahler manifolds the $U(N|M)$ 
extended susy algebra  is not first class, as additional independent operators
appear on the right hand sides, as evident for example in eqs. 
(\ref{QH commutator}) and (\ref{QbarQ anticomm}).
However, there are special  cases, namely the $U(1|0)$ and $U(2|0)$ quantum mechanics, 
which generate first class superalgebras with a central hamiltonian 
on any K\"ahler background. In fact, for the $U(1|0)\equiv U(1)$ model the algebra 
reduces to
\begin{equation}
 \{Q, \bar Q\}=\hbar\,H\;,\quad [Q,H]=0
\end{equation}
where the hamiltonian is now defined by
\begin{equation}
 H=H_0-\frac{\hbar}{2}\,R^\mu_\nu\,M^\nu_\mu+\frac{\hbar^2}{4}\,R=H_0^{sym}+\frac{\hbar^2}{4}\,R\;,
\end{equation}
with $H_0^{sym}=\frac12 g^{\mu\bar\nu}(\pi_\mu\bar\pi_{\bar\nu}+\bar\pi_{\bar\nu}\pi_\mu)$. 
For the $U(2|0)\equiv U(2)$ model the choice of the hamiltonian is the canonical one, \emph{i.e.} the one in \eqref{H with cs} with $c_1=c_2=-\frac12$ and $c_3=0$, and the superalgebra closes as
\begin{equation}
 \{Q_a, \bar Q^b\}=\delta_a^b\,H\;,\quad [Q_a,H]=0\;.
\end{equation}

For the general $U(N|M)$ extended susy algebras one cannot achieve
such generality. Nevertheless, one may look for special backgrounds that
 make \eqref{QH commutator} and \eqref{QbarQ anticomm} first class.
A nontrivial class of K\"ahler manifolds where the first class property
can be achieved is that of manifolds with 
constant holomorphic sectional curvature. On these manifolds, the Riemann and Ricci 
tensors take the form
\begin{equation}
R_{\mu\bar\nu\sigma\bar\lambda} = -\frac{R}{d(d+1)}(g_{\mu \bar\nu} g_{\sigma\bar\lambda}+
 g_{\sigma \bar\nu} g_{\mu\bar\lambda})  \;, \qquad 
 R_{\mu\bar\nu} = \frac{R}{d}g_{\mu \bar\nu} 
 \end{equation}
 where $R$ is the constant scalar curvature. Substituting these relations into the algebra,
 one notices that the metric tensor gets contracted 
with the $Z$ and $\bar Z$ operators, producing additional charges $J_A^B$
on the right hand side, so that with a suitable redefinition of the hamiltonian
one obtains a first class algebra
for generic $m$, $s$, $c_1$ and $c_2$, while $c_3$ 
gets fixed to a unique value.
There is no loss of generality in choosing $c_1$ and $c_2$ equal to their canonical values, 
$c_1=c_2=-\frac{1}{2}$, when using the algebra as a first class constraint algebra. In this case  
\begin{equation}
\begin{split}
 c_3 &= -\frac{m}{2d(d+1)}\Big ((N-M)^2 +(N-M)(4d-3m -2s+1) +2 (m-d)\Big) \\
&+ \frac{s}{2}\Big (1+ \frac{2(d-m)}{d} -\frac{s}{d+1} \Big )
\end{split}
\end{equation}
and the algebra can be casted in the following form
\begin{equation}\label{algebra in constant curvature background}
\begin{split}
[Q_A, \bar Q^B\}&=\hbar\delta_A^B\,H-\frac{\hbar\,R}{d(d+1)}\,\Big\{(-)^{(A+B)C}J_A^CJ_C^B+(-)^{AB}J_A^BJ+(-)^{AB}\hbar k_1J_A^B \\
&+\delta_A^B \Big(\frac12 J^C_D\epsilon^D_E J^E_C+\frac12 J^2+\hbar k_2J\Big) \Big\}\;,\\
[Q_A, H]&=0
\end{split}
\end{equation}
where
\begin{equation}
\begin{split}
k_1 &= d- s(d+1) + m(N-M-2)\\
k_2 &= d-s(d+1) -\Big(m+\frac12 \Big )(N-M) +\frac12\;.
\end{split}
\end{equation}
 We denoted $J\equiv J^A_A$
and used the notation $(-)^{A}$ with $A=0$ for a bosonic index and $A=1$ for a fermionic one.
Gauging this first class algebra produces ``$U(N|M)$ spinning particles" on 
K\"ahler manifolds with constant holomorphic curvature, in a way analogous to
the coupling of standard ``$O(N)$ spinning particles" to (A)dS spaces constructed in 
\cite{Kuzenko:1995mg}.

One may recall that
K\"ahler spaces with constant holomorphic sectional curvature are a subclass of
spaces with vanishing Bochner tensor. The latter is a 
sort of complex analogue of the riemannian Weyl tensor, introduced in \cite{Bochner} and defined by 
\begin{equation}
\begin{split}
 B_{\mu\bar\nu\sigma\bar\lambda} &= 
R_{\mu\bar\nu\sigma\bar\lambda} +\frac{1}{d+2}(  
 g_{\mu \bar\nu} R_{\sigma\bar\lambda}+  g_{\sigma\bar\lambda} R_{\mu\bar\nu}
 +  g_{\sigma \bar\nu} R_{\mu\bar\lambda} +
 g_{\mu\bar\lambda} R_{\sigma \bar\nu} ) \\
& -\frac{R}{(d+1)(d+2)}(g_{\mu \bar\nu} g_{\sigma\bar\lambda}+
 g_{\sigma \bar\nu} g_{\mu\bar\lambda}) \;.
 \end{split}
 \end{equation}
It satisfies the nice property of being traceless, $g^{\mu\bar\nu}B_{\mu\bar\nu\sigma\bar\lambda} =0$.
It seems likely that on spaces with vanishing  Bochner tensor one may obtain a first class algebra, indeed it
is relatively easy to verify it at the classical level, but we do not wish to pursue the detailed quantum analysis here.
 
\section{Transition amplitude}

Up to now we have discussed nonlinear sigma models with
$U(N|M)$ extended supersymmetry, broken at times by the target space geometry, 
and used them to analyze algebraic properties
of differential operators defined on K\"ahler manifolds. 
The aim of this section is the explicit computation of the transition amplitude in euclidean time,
that is $\Braket{x\,\bar\eta}{y\,\xi}{e^{-\frac{\beta}{\hbar} H}}$,
in the limit of short propagation time and using operatorial methods. 
Such a calculation was presented for standard nonlinear sigma models with one, two or no supersymmetries
in \cite{Peeters:1993vu}, see also \cite{Bastianelli:2006rx}, with the main
purpuse of identifying a benchmark to which compare path integral evaluations of the same heat kernel. 
As we wish to be able to master path integrals for $U(N|M)$ sigma models, and eventually use them to address
quantum properties of higher spin equations on K\"ahler manifolds,
we compute here the heat kernel using the operatorial formulation of quantum mechanics.
To achieve sufficient generality and allow diverse applications, we compute the heat kernel for the general
hamiltonian \eqref{H with cs} containing three arbitrary couplings $(c_1, c_2, c_3)$ to the background curvature
plus a fourth one, the charge $s$, hidden in the $U(1)$ part of the connection, see eq. \eqref{s charge}.

Before starting the actual computation, we shall review our set up.
We work on a $2d$ real dimensional K\"{a}hler manifold as target space.
Holomorphic and anti-holomorphic vector indices will be often grouped into
 a riemannian index $i=(\mu,\bar\mu)$ for sake of brevity. The metric in
holomorphic coordinates factorizes as follows
\begin{equation}
g_{ij}=\left(\begin{array}{cc} 0 & g_{\mu\bar\nu} \\ g_{\bar\mu\nu} & 0 \end{array}\right)\;.
\end{equation}
For determinants we use the conventions
$g=\det(g_{\mu\bar\nu})$ and $G=\abs{\det(g_{ij})}=\abs{g}^2$. The dynamical 
variables of the $U(N|M)$
supersymmetric quantum mechanics consist of the following operators:
target space coordinates $(x^\mu,\bar x^{\bar\mu})=x^i$, conjugate momenta $p_i$, and
graded vectors $Z^\mu_A$ and $\bar Z_\nu^A$. 
Their fundamental  (anti)-commutation relations are given in \eqref{CCR}.
For computational advantages we recast the full quantum hamiltonian \eqref{H with cs}
 in a way that directly shows the dependence on the $Z$ operators
\begin{equation}\label{full quantum H}
\begin{split}
H &= H_0+\Delta H\quad\text{with}\\
H_0 &= g^{\bar\mu\nu}\,g^{1/2}\,\bar
p_{\bar\mu}\,\big(p_\nu-i\,\Gamma^\lambda_{\nu\sigma}\,M^\sigma_\lambda\big)\,g^{-1/2}\\
\Delta H &= a_1\,R_{\mu\ph\nu\rho}^{\ph\mu\nu\ph\rho\sigma}\,\bar Z_\nu\cdot Z^\mu\,\bar
Z_\sigma\cdot Z^\rho+a_2\,\hbar\,R^\mu_\nu\,\bar Z_\mu\cdot Z^\nu+a_3\,\hbar^2\,R\;,
\end{split}
\end{equation}
where the $a$ couplings are related to the $c$ couplings by 
\begin{equation}
 a_1=c_1\;,\quad a_2=c_2+2sc_1\;,\quad a_3= c_3-sc_2-s^2c_1 \;.
\end{equation}
Finally, it useful to recall that
the final answer for the heat kernel will contain the exponent of the classical 
action, suitably Wick-rotated to euclidean time $\tau$  ($t\to -i\tau$),
which in phase space takes the form
\begin{equation}
S=\int_{-\beta}^0 d\tau\,\Big[-ip_\mu\dot x^\mu-i\bar p_{\bar\mu}\dot{\bar x}^{\bar\mu}+\bar Z_\mu^A\dot
Z^\mu_A+H_{cl}\Big]
\end{equation}
where $H_{cl}$ is the classical hamiltonian, a function, modified by suitable quantum corrections
depending on $\hbar$.

Now we are ready for the explicit computation of the transition amplitude, through order
$\beta$  (up to the leading free particle propagator), 
between position eigenstates and coherent
states for the internal degrees of freedom, \emph{i.e.}
\begin{equation}\label{transition amplitude definition}
\Braket{x\,\bar\eta}{y\,\xi}{e^{-\frac{\beta}{\hbar} H}}\;,
\end{equation}
where $Z^\mu_A\ket{\xi}=\xi^\mu_A\ket{\xi}$ and $\bra{\bar\eta}\bar
Z_\mu^A=\bra{\bar\eta}\bar\eta_\mu^A$. Of course, $\ket{x}$ and $\ket{y}$ denote  eigenvectors of the position operator $x^i$ as usual,
 $\ket{y\,\xi} \equiv \ket{y}\otimes \ket{\xi}$, and so on.
For convenience in the normalization of the coherent states, from now on we rescale the $Z$ fields by a factor of $\sqrt{\hbar}$, so that $[Z^\mu_A, \bar Z_\nu^B\}=\delta^\nu_\mu\,\delta_A^B$.  We are going to insert in \eqref{transition amplitude definition} a
complete set of momentum eigenstates, and as an intermediate stage we need to compute
\begin{equation}\label{transition x-p}
\Braket{x\,\bar\eta}{p\,\xi}{e^{-\frac{\beta}{\hbar} H}} \;,
\end{equation}
pushing all $p$'s  and $Z$'s to the right, all $x$'s and $\bar Z$'s to the left, taking into
account all (anti)-commutators and then substituting these operators with the corresponding eigenvalues. Let us
focus on the evaluation of \eqref{transition x-p}; clearly we have
\begin{equation}\label{sum}
\Braket{x\,\bar\eta}{p\,\xi}{e^{-\frac{\beta}{\hbar}
H}}=\sum_{k=0}^\infty\frac{(-)^k}{k!}\,\left(\frac{\beta}{\hbar}\right)^k\,\Braket{x\,\bar\eta}{p\,\xi}{H^k}\;.
\end{equation}
It is well known that, in the case of a nonlinear sigma model, it is not sufficient to expand the
exponent to first order, \emph{i.e.} $e^{-\beta H/\hbar}\sim 1-\frac{\beta}{\hbar}H$, to obtain the
correct transition amplitude to order $\beta$, see \cite{Peeters:1993vu,Bastianelli:2006rx}.
Contributions for all $k$ must
be retained in the sum \eqref{sum}, but taking into account at most two $[x,p]$ commutators. Let us
see this in more detail. In a factor of $H^k$, pushing all $p$'s to the right by repeated use of
the $[x,p]$ commutator, one obtains, remembering that each $H$ can give at most two $p$
eigenvalues,
\begin{equation}\label{Bcoeff}
\Braket{x\,\bar\eta}{p\,\xi}{H^k}=\sum_{l=0}^{2k}B^k_l(x,\bar\eta,\xi)\,p^l\,\braket{x\,\bar\eta}{p\,\xi}\;,
\end{equation}
where $p^l$ stands for a homogeneous polynomial in $p$ of degree $l$. For the position eigenstates
we use the normalization: $\braket{x}{x'}=g^{-1/2}(x)\delta^{2d}(x-x')$, while the standard
normalization is employed for $p$-eigenstates. In this way the completeness relations read
\begin{equation}
\mathbf1=\int d^{2d}p\,\ket{p}\bra{p}\quad,\quad\mathbf1=\int d^{2d}x\,g\,\ket{x}\bra{x}\;,
\end{equation}
while the plane waves are given by: $\braket{x}{p}=(2\pi\hbar)^{-d}g^{-1/2}(x)e^{ip\cdot x}$, with
$p\cdot x\equiv p_ix^i=p_\mu x^\mu+\bar p_{\bar\mu}\bar x^{\bar\mu}$. Finally, coherent states are
normalized as $\braket{\bar\eta}{\xi}=e^{\bar\eta\cdot\xi}$. Having set our normalizations, we 
expand the transition amplitude as follows
\begin{equation}
\begin{split}
&\Braket{x\,\bar\eta}{y\,\xi}{e^{-\frac{\beta}{\hbar} H}}=(2\pi\hbar)^{-d}\,g^{-1/2}(y) \int
d^{2d}p\,e^{-\frac{i}{\hbar}p\cdot y}\,\Braket{x,\bar\eta}{p\,\xi}{e^{-\beta H/\hbar}}\\
&= (2\pi\hbar)^{-2d}\,[g(x)g(y)]^{-1/2}\int
d^{2d}p\,e^{\frac{i}{\hbar}p\cdot(x-y)}\,e^{\bar\eta_\mu\cdot\xi^\mu}\,\sum_{k=0}^\infty\left(-\frac{\beta}{\hbar}\right)^k\frac{1}{k!}
\sum_{l=0}^{2k}B_l^k(x,\bar\eta,\xi)\,p^l\;.
\end{split}
\end{equation}
Now, to make the $\beta$ dependence explicit, we rescale momenta as $p_i=\sqrt{\hbar/\beta}q_i$ and
obtain
\begin{equation}\label{beta dependence}
\begin{split}
\Braket{x\,\bar\eta}{y\,\xi}{e^{-\frac{\beta}{\hbar} H}} &=
(4\pi^2\hbar\beta)^{-d}[g(x)g(y)]^{-1/2}e^{\bar\eta_\mu\cdot\xi^\mu}\int
d^{2d}q\,e^{iq\cdot(x-y)/\sqrt{\beta\hbar}} \\
&\times\sum_{k=0}^\infty\frac{(-)^k}{k!}
\sum_{l=0}^{2k}\left(\frac{\beta}{\hbar}\right)^{k-l/2}B^k_l(x,\bar\eta,\xi)\,q^l\;.
\end{split}
\end{equation}
After momentum integration, in configuration space the leading term in $(x-y)$ will be of the form
$\exp[-(x-y)^2/2\beta\hbar]$, showing that effectively $(x-y)\sim\mathcal{O}(\beta^{1/2})$. Then,
looking at \eqref{beta dependence}, we see that $q\sim\mathcal{O}(\beta^0)$ and so in the sum over
$l$ only $B^k_{2k}$, $B^k_{2k-1}$ and $B^k_{2k-2}$ will contribute, for all $k$, to the order
$\beta$ amplitude, as anticipated\footnote{Note that in $B_l^k$ at most $2k-l$ $[x,p]$
commutators are taken into account.}.

The $B^k_l$ coefficients are explicitly derived in appendix \ref{computation of B coefficients},
and inserting \eqref{B2k-1} and \eqref{B2k-2} into \eqref{beta
dependence}, one can see that the sum in $k$ can be immediately performed, producing the gaussian
exponential $\exp[-q^2/2]$. The transition amplitude \eqref{beta dependence} then becomes
\begin{equation}\label{transition amplitude with momenta}
\begin{split}
&\Braket{x\,\bar\eta}{y\,\xi}{e^{-\frac{\beta}{\hbar} H}}=
(4\pi^2\hbar\beta)^{-d}[g(x)g(y)]^{-1/2}e^{\bar\eta_\mu\cdot\xi^\mu}\int
d^{2d}q\,e^{-q^2/2-iq\cdot\Delta/\sqrt{\beta\hbar}}\,\Big\{1+\sqrt{\beta\hbar}\,\Big[
\frac{i}{2}g^jq_j\\
&-\frac{i}{4}g^{klj}\,q_j\,q_k\,q_l+ig^{\bar\mu\nu}\,\Gamma^\lambda_{\nu\sigma}\,
(\bar\eta_\lambda\cdot\xi^\sigma)'\bar q_{\bar\mu}\Big]+\beta\hbar\,\Big[-\frac{1}{32}\ln G_i\ln
G^i-\frac18\ln G_i^i-\frac18g^i\ln
G_i\\
&-\Big(\frac14\de^jg^l+\frac18g^jg^l+\frac18g^kg^{jl}_k+\frac18g^{jlk}_k\Big)\,q_j\,q_l+\Big(\frac{1}{12}
g^{mnkl}+\frac18g^{klm}g^n+\frac{1}{12}g^{ikl}g^{mn}_i\\
&+\frac{1}{24}g^{kl}_ig^{mni}\Big)\,q_k\,q_l\,q_m\,q_n-\Big(\frac{1}{32}g^{klj}g^{pqm}\Big)\,q_j\,q_k\,q_l\,
q_m\,q_p\,q_q-\frac12g^{ij}\de_j\Big(g^{\mu\bar\nu}\,\Gamma^\lambda_{\mu\sigma}\Big)
(\bar\eta_\lambda\cdot\xi^\sigma)'q_i\,\bar q_{\bar\nu}\\
&-\frac12g^{\bar\mu\nu}\,\Gamma^\rho_{\nu\sigma}\,
(\bar\eta_\rho\cdot\xi^\sigma)'\Big(\de_{\bar\mu}g^{\lambda\bar\sigma}\,q_\lambda\,\bar
q_{\bar\sigma}+g^jq_j\,\bar q_{\bar\mu}-\frac12g^{klj}\,q_j\,q_k\,q_l\,\bar
q_{\bar\mu}+g^{\lambda\bar\sigma}\de_{\bar\mu}g_{\lambda\bar\sigma}\Big)\\
&-a_1\,R^{\ph\mu\nu\ph\rho\sigma}_{\mu\ph\nu\rho}
\,\bar\eta_\nu\cdot\xi^\mu\bar\eta_\sigma\cdot\xi^\rho-(a_2-a_1+1)\,R^\mu_\nu\,\bar\eta_\mu\cdot\xi^\nu-
\Big(a_3-s\Big)\,R\\
&-\frac12g^{\bar\mu\nu}\Gamma^\mu_{\nu\tau}g^{\lambda\bar\sigma}\Gamma^\rho_{\lambda\sigma}
\,\bar q_{\bar\mu}\,\bar
q_{\bar\sigma}\,\big[(\bar\eta_\mu\cdot\xi^\tau)'(\bar\eta_\rho\cdot\xi^\sigma)'+\delta_\rho^\tau\,\bar\eta_\mu\cdot
\xi^\sigma\big]\Big]\Big\}\;,
\end{split}
\end{equation}
where $\Delta^i=y^i-x^i$ and $(\bar\eta_\lambda\cdot\xi^\sigma)'=(\bar\eta_\lambda\cdot\xi^\sigma-s\,\delta_\lambda^\sigma)$.
In order to lighten the formulae we have used the following compact notation
\begin{equation}\nonumber
\begin{split}
\de_i...\de_mg^{jk}  &= g^{jk}_{i...m}\;,\quad g^{ij}g^{kl}_j=g^{kli}\;,\quad g^{ij}_j=g^i \\
 g^{jk}\de_kg^{lm}_m
&=\de^jg^l\;,\quad\de_i\ln G=\ln G_i\;,\quad g^{ij}\de_i\de_j\ln G=\ln G_i^i \;. 
\end{split}
\end{equation}
Now we can complete squares in the exponent of \eqref{transition
amplitude with momenta}, shift integration variables and perform the gaussian integral over
momenta. The transition amplitude, up to order $\beta$, is then given by
\begin{equation}\label{transition amplitude expanded}
\begin{split}
&\Braket{x\,\bar\eta}{y\,\xi}{e^{-\frac{\beta}{\hbar}
H}}=(2\pi\hbar\beta)^{-d}\,\big[g(x)/g(y)\big]^{1/2}\,e^{-\frac{1}{2\beta\hbar}g_{ij}\Delta^i\Delta^j}\,
e^{\bar\eta_\mu\cdot\xi^\mu}\Big\{1+\Delta^i\,g^{-1/2}\,\de_i\,g^{1/2}\\
&-\frac{1}{4\beta\hbar}\,\de_kg_{ij}\,\Delta^i\Delta^j\Delta^k+\frac12\,\Delta^i\Delta^j\,g^{-1/2}\,\de_i\de_jg^{1/2}
-\frac{1}{4\beta\hbar}\,\Delta^ig^{-1/2}\de_ig^{1/2}\de_kg_{mn}\,\Delta^k\Delta^m\Delta^n\\
&+\frac12\,\Big[\frac{1}{4\beta\hbar}\,\de_kg_{ij}\,\Delta^i\Delta^j\Delta^k\Big]^2-\frac{1}{12\beta\hbar}
\,\Big[\de_k\de_lg_{ij}-\frac12g_{mn}\,\Gamma^m_{ij}\,\Gamma^n_{kl}\Big]\Delta^i\Delta^j\Delta^k\Delta^l+\frac{1}{6}\,
R_{\mu\bar\nu}\,\Delta^\mu\bar\Delta^{\bar\nu}\\
&+\Delta^\nu\,\Gamma^\lambda_{\nu\sigma}\,(\bar\eta_\lambda\cdot\xi^\sigma)'+
\Big[\Delta^\nu\Gamma^\lambda_{\nu\sigma}\,(\bar\eta_\lambda\cdot\xi^\sigma)'\Big]\Big[\Delta^ig^{-1/2}\de_ig^{1/2}\Big]
+\frac12\Big[\Delta^\nu\Gamma^\lambda_{\nu\sigma}\,(\bar\eta_\lambda\cdot\xi^\sigma)'\Big]^2\\
&-\frac{1}{4\beta\hbar}
\,\de_jg_{kl}\,\Delta^j\Delta^k\Delta^l\,\Big(\Delta^\nu\Gamma^\mu_{\nu\sigma}\,(\bar\eta_\mu\cdot\xi^\sigma)'\Big)
+\frac12\Delta^i\Delta^\mu\de_i\Gamma^\lambda_{\mu\sigma}\,(\bar\eta_\lambda\cdot\xi^\sigma)'\\
&+\frac12\,\Delta^\nu\Delta^\lambda\,\Gamma^\mu_{\nu\sigma}\Gamma^\sigma_{\lambda\rho}\,\bar\eta_\mu\cdot\xi^\rho
-a_1\,\beta\hbar\,R_{\mu\ph\nu\rho}^{\ph\mu\nu\ph\rho\sigma}
\,\bar\eta_\nu\cdot\xi^\mu\,\bar\eta_\sigma\cdot\xi^\rho+\Big(a_1-a_2-\frac12\Big)\,\beta\hbar\,R^\mu_\nu\,\bar\eta_\mu\cdot\xi^\nu\\
&+\Big(\frac{1}{6}+\frac{s}{2}-a_3\Big)\,\beta\hbar\,R+\mathcal{O}(\beta^{3/2})\Big\}\;.
\end{split}
\end{equation}
All functions in \eqref{transition amplitude expanded}, if not specified otherwise, are
evaluated at point $x$. Keeping in mind that the transition amplitude is a bi-scalar, and that
in a semiclassical expansion the classical action evaluated on-shell should appear in the exponent, 
we factorize and exponentiate, up to order $\beta$, four terms
\begin{equation}\label{transition amplitude factorized}
\begin{split}
&\Braket{x\,\bar\eta}{y\,\xi}{e^{-\frac{\beta}{\hbar}
H}}=(2\pi\hbar\beta)^{-d}\,g(y)^{-1/2}\Big[g^{1/2}+\Delta^i\de_ig^{1/2}+\frac12\,\Delta^i\Delta^j\de_i\de_jg^{1/2}\Big]\\
&\exp\Big\{\!\!-\frac{1}{\beta\hbar}\,\Big[\frac12\,g_{ij}\,\Delta^i\Delta^j+\frac14\,\de_ig_{jk}\,\Delta^i\Delta^j\Delta^k+\frac{1}{12}\,
\Big(\de_k\de_lg_{mn}-\frac12\,g_{ij}\,\Gamma^i_{kl}\,\Gamma^j_{mn}\Big)\,\Delta^k\Delta^l\Delta^m\Delta^n\Big]\Big\}\\
&\exp\Big\{\bar\eta_\mu\cdot\xi^\mu+\Delta^\nu\,\Gamma^\lambda_{\nu\sigma}\,(\bar\eta_\lambda\cdot\xi^\sigma)'
+\frac12\Delta^i\Delta^\mu\de_i\Gamma^\lambda_{\mu\sigma}\,(\bar\eta_\lambda\cdot\xi^\sigma)'
+\frac12\,\Delta^\nu\Delta^\lambda\,\Gamma^\mu_{\nu\sigma}\Gamma^\sigma_{\lambda\rho}\,\bar\eta_\mu\cdot\xi^\rho\\
&-a_1\,\beta\hbar\,R_{\mu\ph\nu\rho}^{\ph\mu\nu\ph\rho\sigma}
\,\bar\eta_\nu\cdot\xi^\mu\,\bar\eta_\sigma\cdot\xi^\rho-a_2\,\beta\hbar\,R^\mu_\nu\,\bar\eta_\mu\cdot\xi^\nu
-a_3\,\beta\hbar\,R\Big\}\\
&\Big[1+\frac16\,R_{\mu\bar\nu}\,\Delta^\mu\bar\Delta^{\bar\nu}+\Big(a_1-\frac12\Big)\,\beta\hbar\,
R^\mu_{\nu}\,\bar\eta_\mu\cdot\xi^\nu+\Big(\frac16+\frac{s}{2}\Big)\,\beta\hbar\,R\Big]\;.
\end{split}
\end{equation}
The first term contains the Taylor expansion around $x$ of $g(y)^{1/2}$, that cancel the $g(y)^{-1/2}$ factor.
The second and third terms should be the expansions of the exponential of the classical action, and the
fourth is evidently covariant. The detailed study of the expansion of the on-shell action is demanded to appendix \ref{classical action appendix}. 
Comparing the result \eqref{classical action expanded} for the classical 
on-shell action $\tilde S_{os}$ with the expansion \eqref{transition amplitude factorized},
we see that, as expected, the transition amplitude can finally be cast in an explicitly covariant form
\begin{equation}\label{transition amplitude covariant}
\begin{split}
\Braket{x\,\bar\eta}{y\,\xi}{e^{-\frac{\beta}{\hbar}
H}}&=(2\pi\hbar\beta)^{-d}\,e^{-\tilde S_{os}/\hbar}\,\Big[1+\frac16\,R_{\mu\bar\nu}\,\Delta^\mu\bar\Delta^{\bar\nu}+\Big(a_1-\frac12\Big)\,\beta\hbar\,
R^\mu_{\nu}\,\bar\eta_\mu\cdot\xi^\nu\\
&+\Big(\frac16+\frac{s}{2}\Big)\,\beta\hbar\,R + {\cal O}(\beta^2)\Big]
\end{split}
\end{equation}
where the coordinate displacements $\Delta^\mu$ are considered of order $\sqrt{\beta}$.
 
\section{Conclusions and outlook}

In this paper we have introduced and studied the quantum properties of 
a class of quantum mechanical models with $U(N|M)$ extended supersymmetry on the worldline.
These models take the form of nonlinear sigma models with K\"ahler manifolds as target spaces,
and can be interpreted as describing the motion of a particle with extra degrees of freedom,
carried by graded complex vectors $Z^\mu_A$, on K\"ahler spaces.
When the K\"ahler space is flat, the model has conserved charges satisfying precisely a  
$U(N|M)$ extended supersymmetry algebra on the worldline.
On curved K\"ahler spaces, the charges get modified by the geometry  as does the corresponding 
quantum algebra, which generically fails to be first class, 
though a symmetry under the supergroup $U(N|M)$  is always present. 
Conserved supercharges can be defined on locally symmetric K\"ahler manifolds, 
i.e. K\"ahler manifolds with covariantly constant curvature tensors, while  
a truly first class algebra can be obtained on K\"ahler manifolds with constant holomorphic sectional curvature.
The latter case is particularly interesting, as one can gauge the symmetry  
charges to obtain higher spin equations with peculiar gauge symmetries,
as studied in flat space for the $U(N|0)$ models in  \cite{Bastianelli:2009vj}. 

In the second part of the paper we have computed the heat kernel 
for our quantum mechanical models in a perturbative expansion.
The computation was performed with operatorial methods
on arbitrary K\"ahler manifolds and with a
general hamiltonian containing four arbitrary couplings.
The calculation turned out to be somewhat tedious
for a rather simple final result. One possible application of this result is to use it as
a benchmark for path integral calculations, which are often simpler 
and more flexible, but need to be defined precisely, with predetermined regularization schemes and 
corresponding counterterms. Indeed the operatorial calculation of ref.
\cite{Peeters:1993vu} was useful to identify the correct time slicing regularization
of path integrals in curved spaces \cite{DeBoer:1995hv}.
Correctness of  the alternative but equivalent mode \cite{Bastianelli:1991be}
and dimensional  \cite{Kleinert:1999aq, Ba:2005vk} regularizations
has then been checked against time slicing, and the full consistency of
these three schemes have been instrumental in putting the method of 
path integration on curved manifolds on solid foundations  \cite{Bastianelli:2006rx}. 
In future works we plan to construct regularized path integrals for the $U(N|M)$ quantum mechanics,
use them to study effective actions induced by higher spin fields and compute higher order heat kernel coefficients.

\acknowledgments{We wish to thank Andrew Waldron for discussions. 
This work was supported in part by the Italian MIUR-PRIN contract 20075ATT78.}
\vskip 1cm

\appendix

\section{Computation of the $B^k_l$ coefficients}\label{computation of B coefficients}

In order to compute the $B$ coefficients defined in eq. \eqref{Bcoeff}
we follow the strategy explained in  \cite{Peeters:1993vu},
and divide the
hamiltonian \eqref{full quantum H} in three pieces, contributing at most two, one or no $p$
eigenvalues, respectively\footnote{Remember that we are using rescaled $Z$'s.}
\begin{equation}
\begin{split}
H &= H_B+H_1+H_2\qquad\text{where}\\
H_B &= g^{\bar\mu\nu}\,g^{1/2}\,\bar
p_{\bar\mu}\,p_\nu\,g^{-1/2} = \frac12\,G^{-1/4}\,p_i\,G^{1/2}\,g^{ij}\,p_j\,G^{-1/4}\;,\\
H_1 &= -i\hbar\,g^{\bar\mu\nu}\,\Gamma^\lambda_{\nu\sigma}\,\left(\bar Z_\lambda\cdot
Z^\sigma-s\,\delta^\sigma_\lambda\right)\,
g^{1/2}\,\bar p_{\bar\mu}\,g^{-1/2}\;,\\
H_2 &= a_1\hbar^2\,R^{\ph\mu\nu\ph\rho\sigma}_{\mu\ph\nu\rho}\,\bar Z_\nu\cdot Z^\mu\bar
Z_\sigma\cdot Z^\rho+(a_2+1)\hbar^2\,R^\nu_\mu\,\bar Z_\nu\cdot
Z^\mu+\left(a_3-s\right)\hbar^2\,R\;.
\end{split}
\end{equation}
First of all, notice that $H_B$ is precisely the usual bosonic quantum hamiltonian, carefully studied
in the literature \cite{Peeters:1993vu, Bastianelli:2006rx}.
 Let us start with $B^k_{2k}$: the only way to have $2k$
$p$ eigenvalues is $k$ factors of $H_B$ and no commutators taken into account, giving simply
\begin{equation}
B^k_{2k}\,p^{2k}=\left(\frac{p^2}{2}\right)^k\;,
\end{equation}
where we use the notation $p^2=g^{ij}p_ip_j=2g^{\mu\bar\nu}p_\mu\bar p_{\bar\nu}$. For $B^k_{2k-1}$
we can have two terms. The first term comes from $k$ factors of $H_B$ with one $p$ acting as a derivative; this gives the
corresponding $B^k_{2k-1}$ coefficient, that we call $A^k_{2k-1}$, of the purely bosonic model,
whose computation is explained in detail in \cite{Peeters:1993vu, Bastianelli:2006rx}.
The other term comes from $k-1$
factors of $H_B$ and one $H_1$, by substituting all operators with the corresponding eigenvalues.
Putting things together we obtain
\begin{equation}\label{B2k-1}
\begin{split}
B^k_{2k-1}\,p^{2k-1} &= A^k_{2k-1}\,p^{2k-1}-i\hbar
k\,\left(\frac{p^2}{2}\right)^{k-1}\Gamma^\lambda_{\nu\sigma}\,\big(\bar\eta_\lambda\cdot\xi^\sigma\big)'\,g^{\bar\mu\nu}\,\bar
p_{\bar\mu}\\
&= -\frac{i\hbar
k}{2}\,\left(\frac{p^2}{2}\right)^{k-1}g^j\,p_j-i\hbar\,\binom{k}{2}\,\left(\frac{p^2}{2}\right)^{k-2}\,\frac12
g^{klj}\,p_j\,p_k\,p_l\\
&-i\hbar
k\,\left(\frac{p^2}{2}\right)^{k-1}\Gamma^\lambda_{\nu\sigma}\,\big(\bar\eta_\lambda\cdot\xi^\sigma\big)'\,g^{\bar\mu\nu}\,\bar
p_{\bar\mu}\;,
\end{split}
\end{equation}
where we denoted
$(\bar\eta_\lambda\cdot\xi^\sigma)'=(\bar\eta_\lambda\cdot\xi^\sigma-s\,\delta_\lambda^\sigma)$,
$g^j=\de_ig^{ij}$ and $g^{ijk}=g^{kl}\,\de_lg^{ij}$. For $B^k_{2k-2}$ four types of term
contribute: \emph{i)} $k$ factors of $H_B$, giving the corresponding coefficient $A^k_{2k-2}$,
\emph{ii)} $k-1$ factors of $H_B$ and one $H_1$, with one $p$ acting as a derivative. This
contribution gives four terms: the derivative acting from one $H_B$ to $H_1$, from $H_1$ to one
$H_B$, within $H_1$ or within the $k-1$ $H_B$'s. \emph{iii)} $k-1 $ factors of $H_B$ and one $H_2$,
substituting all operators with their eigenvalues, and \emph{iv)} $k-2$ factors of $H_B$ and two
$H_1$, substituting all with eigenvalues. Remember that in \emph{iii)} and \emph{iv)} $[Z,\bar Z\}$
(anti)-commutators have to be taken into account in order to obtain eigenvalues on the coherent
states. Altogether it results in
\begin{equation}\label{B2k-2}
\begin{split}
&B^k_{2k-2}\,p^{2k-2} =
A^k_{2k-2}\,p^{2k-2}-\hbar^2\binom{k}{2}\left(\frac{p^2}{2}\right)^{k-2}g^{ij}\de_j\Big(g^{\bar\mu\nu}\Gamma^\lambda_{\nu\sigma}\Big)
\big(\bar\eta_\lambda\cdot\xi^\sigma\big)'p_i\,\bar p_{\bar\mu}\\
&-\hbar^2\binom{k}{2}\left(\frac{p^2}{2}\right)^{k-2}\!\!\!g^{\bar\mu\nu}\Gamma^\lambda_{\nu\sigma}
\big(\bar\eta_\lambda\cdot\xi^\sigma\big)'\de_{\bar\mu}g^{\rho\bar\sigma}\,p_\rho\,\bar
p_{\bar\sigma}+\frac12\,\hbar^2k\left(\frac{p^2}{2}\right)^{k-1}\!\!\!g^{\bar\mu\nu}\Gamma^\lambda_{\nu\sigma}
\big(\bar\eta_\lambda\cdot\xi^\sigma\big)'g^{\rho\bar\sigma}\de_{\bar\mu}g_{\rho\bar\sigma}\\
&-i\hbar k\,A^{k-1}_{2k-3}\,p^{2k-3}\,g^{\bar\mu\nu}\Gamma^\lambda_{\nu\sigma}
\big(\bar\eta_\lambda\cdot\xi^\sigma\big)'\bar
p_{\bar\mu}+\hbar^2k\left(\frac{p^2}{2}\right)^{k-1}\Big(a_1\,R^{\ph\mu\nu\ph\rho\sigma}_{\mu\ph\nu\rho}\,
\bar\eta_\nu\cdot\xi^\mu\bar\eta_\sigma\cdot\xi^\rho\\
&+(a_2-a_1+1)\,R^\mu_\nu\,\bar\eta_\mu\cdot\xi^\nu+\Big(a_3-s\Big)\,R\Big)-\hbar^2\binom{k}{2}\left(\frac{p^2}{2}\right)^{k-2}
g^{\bar\mu\nu}\Gamma^\tau_{\nu\sigma}\,g^{\lambda\bar\sigma}\Gamma^\rho_{\lambda\mu}\,\bar
p_{\bar\mu}\,\bar p_{\bar\sigma}\\
&\times\Big[\big(\bar\eta_\tau\cdot\xi^\sigma\big)'\big(\bar\eta_\rho\cdot\xi^\mu\big)'+\delta_\rho^\sigma\,
\bar\eta_\tau\cdot\xi^\mu\Big]\;.
\end{split}
\end{equation}
In the formulae above the bosonic coefficients are given by
\begin{equation}\label{A2k-2}
\begin{split}
A^{k-1}_{2k-3} &=
-\frac{i\hbar}{2}\,(k-1)\left(\frac{p^2}{2}\right)^{k-2}g^jp_j-i\hbar\binom{k-1}{2}\left(\frac{p^2}{2}\right)^{k-3}
\frac12g^{klj}\,p_j\,p_k\,p_l\;,\\
A^k_{2k-2} &= \hbar^2k\left(\frac{p^2}{2}\right)^{k-1}\Big[\frac{1}{32}\ln G_i\ln G^i+\frac18\ln
G_i^i+\frac18g^j\ln G_j\Big]-\hbar^2\binom{k}{2}\left(\frac{p^2}{2}\right)^{k-2}\\
&\times \Big[\frac12\de^jg^l+\frac14g^jg^l+\frac14g^kg^{jl}_k+\frac14g^{jlk}_k\Big]\,p_j\,p_l
-\hbar^2\binom{k}{3}\left(\frac{p^2}{2}\right)^{k-3}\Big[\frac12g^{mnkl}+\frac34g^{klm}g^n\\
&+\frac12g^{ikl}g^{mn}_i+\frac14g^{kl}_ig^{mni}\Big]\,p_k\,p_l\,p_m\,p_n-\hbar^2\binom{k}{4}\left(\frac{p^2}{2}\right)^{k-4}
\Big[\frac34g^{klj}g^{pqm}\Big]\,p_j\,p_k\,p_l\,p_m\,p_p\,p_q
\end{split}
\end{equation}
and we recall that the following compact notation was employed
\begin{equation}\nonumber
\begin{split}
\de_i...\de_mg^{jk} &= g^{jk}_{i...m}\;,\quad g^{ij}g^{kl}_j=g^{kli}\;,\quad g^{ij}_j=g^i\\
g^{jk}\de_kg^{lm}_m&=\de^jg^l\;,\quad\de_i\ln G=\ln G_i\;,\quad g^{ij}\de_i\de_j\ln G=\ln G_i^i \ .
\end{split}
\end{equation}

\section{The on-shell action}\label{classical action appendix}

The euclidean action generated by the hamiltonian \eqref{full quantum H} is given by\footnote{Remember that we are using rescaled $Z$'s.}
\begin{equation}\label{classical action}
S=\int_{-\beta}^0d\tau\,\Big[g_{\mu\bar\nu}\,\dot x^\mu\dot{\bar x}^{\bar\nu}+\hbar\,\bar Z_\mu\cdot\frac{DZ^\mu}{D\tau}
-s\hbar\,\Gamma_\mu\dot x^\mu+\hbar^2\,\Delta H\Big]\;,
\end{equation}
where $\Gamma_\mu\equiv\Gamma^\nu_{\nu\mu}$ is the $U(1)$ piece of the K\"{a}hler connection and $s$ plays the role of an
additional $U(1)$ coupling. The additional piece
\begin{equation}
\Delta H=a_1\,R^{\ph\mu\nu\ph\rho\sigma}_{\mu\ph\nu\rho}\,\bar Z_\nu\cdot Z^\mu\bar Z_\sigma\cdot Z^\rho
+a_2\,R^\mu_\nu\,\bar Z_\mu\cdot Z^\nu+a_3\,R
\end{equation}
contains the generalized couplings to curvatures, and the covariant time derivative on $Z$ fields reads
\begin{equation}
\frac{DZ^\mu_A}{D\tau}=\dot Z^\mu_A+\dot x^\nu\,\Gamma^\mu_{\nu\sigma}\,Z^\sigma_A\;, \qquad
\frac{D\bar Z_\sigma^A}{D\tau}=\dot{\bar Z}_\sigma^A-\dot x^\nu\,\Gamma^\mu_{\nu\sigma}\,Z_\mu^A\;.
\end{equation}
From the action \eqref{classical action} the following equations of motion arise
\begin{equation}\label{eoms}
\begin{split}
& \ddot x^i+\Gamma^i_{jk}\,\dot x^j\dot x^k-\hbar\,R^{i\ph j\mu}_{\ph i j\ph\mu\nu}\,\dot x^j\,\Big(\bar Z_\mu\cdot
Z^\nu-s\,\delta^\nu_\mu\Big)-\hbar^2\,g^{ik}\de_k\Delta H=0\\
& \dot Z^\mu_A+\Gamma^\mu_{\nu\sigma}\,\dot x^\nu\,Z^\sigma_A+2a_1\hbar\,R^{\ph\nu\mu\ph\lambda\sigma}_{\nu\ph\mu\lambda}\,
Z_A^\nu\bar Z_\sigma\cdot Z^\lambda+a_2\hbar\,R^\mu_\nu\,Z^\nu_A=0\\
&\dot{\bar Z}_\nu^A-\Gamma^\mu_{\nu\sigma}\,\dot x^\sigma\,Z_\mu^A-2a_1\hbar\,R^{\ph\nu\mu\ph\lambda\sigma}_{\nu\ph\mu\lambda}\,
\bar Z^A_\mu\bar Z_\sigma\cdot Z^\lambda-a_2\hbar\,R^\mu_\nu\,Z_\mu^A=0\;.
\end{split}
\end{equation}
Now, we have to expand the action \eqref{classical action} up to order $\beta$, with the fields obeying \eqref{eoms}, with
boundary conditions: $x^i(-\beta)=y^i$, $x^i(0)=x^i$, $Z_A^\mu(-\beta)=\xi^\mu_A$ and $\bar Z_\mu^A(0)=\bar\eta_\mu^A$.
Expanding fields in a Taylor series around $\tau=0$, we will see that, for small $\beta$, we have
\begin{equation}\nonumber
\frac{d^nx^i}{d\tau^n}\sim\frac{d^nZ^\mu_A}{d\tau^n}\sim\beta^{-n/2}\;.
\end{equation}
Expanding also the on-shell lagrangian, we can write
\begin{equation}
S_{os}=\sum_{n=0}^\infty\frac{1}{(n+1)!}\,(-)^n\,\left.\frac{d^n\mathcal{L}_{os}}{d\tau^n}\right\rvert_{\tau=0}\,\beta^{n+1}\;,
\end{equation}
and one notices that, for all pieces of the lagrangian but the $U(1)$ one, it is sufficient to keep the order zero: $S_{os}=\beta\,\mathcal{L}_{os}(0)$.
For the $U(1)$ piece, it is necessary the next order: $\beta\,\mathcal{L}_{os}(0)-\frac12\beta^2\dot{\mathcal{L}}_{os}(0)$.
Let us begin with the $x$'s: we expand in Taylor series and obtain
\begin{equation}
x^i(\tau)=\sum_{n=0}^\infty\frac{\tau^n}{n!}\,\frac{d^nx^i}{d\tau^n}(0)\;,
\end{equation}
setting $\tau=-\beta$ and using the boundary conditions we have
\begin{equation}
y^i=x^i-\beta\,\dot x^i(0)+\frac{\beta^2}{2}\,\ddot x^i(0)-\frac{\beta^3}{6}\,\dddot x^i(0)+...
\end{equation}
and similarly for the first derivative
\begin{equation}
\dot x^i(0)=-\frac{\Delta^i}{\beta}+\frac{\beta}{2}\,\ddot x^i(0)-\frac{\beta^2}{6}\,\frac{d}{d\tau}\,\ddot x^i(0) +...\;.
\end{equation}
Now one uses the equations of motion \eqref{eoms} and solves by iteration. To order $\beta^{1/2}$
we obtain
\begin{equation}
\dot x^i(0)=-\frac{\Delta^i}{\beta}-\frac{1}{2\beta}\,\Gamma^i_{jk}\,\Delta^j\Delta^k-\frac{1}{6\beta}\,\Big(
\de_l\Gamma^i_{jk}+\Gamma^i_{js}\,\Gamma^s_{kl}\Big)\,\Delta^j\Delta^k\Delta^l-\frac{\Delta^j}{2}\,\hbar\,
R^{i\ph j\mu}_{\ph i j\ph\mu\nu}\,(\bar\eta_\mu\cdot\xi^\nu)'\;.
\end{equation}
Adopting the same procedure for $Z$ and $\dot Z$ we have
\begin{equation}
\begin{split}
Z^\mu_A(0) &= \xi^\mu_A+\Gamma^\mu_{\nu\lambda}\,\Delta^\nu\xi^\lambda_A\;,\\
\dot Z^\mu_A(0) &= \frac{Z^\mu_A(0)-\xi^\mu_A}{\beta}-\frac{1}{2\beta}\,\de_j\Gamma^\mu_{\nu\lambda}\,\Delta^j\Delta^\nu
\xi^\lambda_A+\frac{1}{2\beta}\,\Gamma^\mu_{\nu\lambda}\,\Gamma^\nu_{\sigma\rho}\,\Delta^\sigma\Delta^\rho\xi^\lambda_A
+\frac{1}{2\beta}\,\Gamma^\mu_{\nu\lambda}\,\Gamma^\nu_{\sigma\rho}\,\Delta^\sigma\Delta^\lambda\xi^\rho_A\;.
\end{split}
\end{equation}
Now we can substitute the above expansions in $\beta\,\mathcal{L}_{os}(0)$ and in $-\frac{\beta^2}{2}\dot{\mathcal{L}}^{U(1)}_{os}
(0)$. Remembering  that in fermionic actions one needs also a boundary term, it is convenient to use the modified
action $\tilde S=S-\hbar\bar Z_\mu(0)\cdot Z^\mu(0)$, and using
\eqref{classical action}  for $S$ we finally arrive at the following expansion
\begin{equation}\label{classical action expanded}
\begin{split}
\tilde S_{os}&=\frac{1}{2\beta}\,g_{ij}\,\Delta^i\Delta^j+\frac{1}{4\beta}\,\de_ig_{jk}\,\Delta^i\Delta^j\Delta^k+
\frac{1}{12\beta}\,
\Big(\de_k\de_lg_{mn}-\frac12\,g_{ij}\,\Gamma^i_{kl}\,\Gamma^j_{mn}\Big)\,\Delta^k\Delta^l\Delta^m\Delta^n\\
&-\hbar\,
\bar\eta_\mu\cdot\xi^\mu-\hbar\,\Delta^\nu\,\Gamma^\lambda_{\nu\sigma}\,(\bar\eta_\lambda\cdot\xi^\sigma)'
-\hbar\,\frac12\Delta^i\Delta^\mu\de_i\Gamma^\lambda_{\mu\sigma}\,(\bar\eta_\lambda\cdot\xi^\sigma)'
-\hbar\,\frac12\,\Delta^\nu\Delta^\lambda\,\Gamma^\mu_{\nu\sigma}\Gamma^\sigma_{\lambda\rho}\,\bar\eta_\mu\cdot\xi^\rho\\
&+a_1\,\beta\hbar^2\,R_{\mu\ph\nu\rho}^{\ph\mu\nu\ph\rho\sigma}
\,\bar\eta_\nu\cdot\xi^\mu\,\bar\eta_\sigma\cdot\xi^\rho+a_2\,\beta\hbar^2\,R^\mu_\nu\,\bar\eta_\mu\cdot\xi^\nu
+a_3\,\beta\hbar^2\,R +...
\end{split}
\end{equation}
which appears in the final result \eqref{transition amplitude covariant}.

\end{document}